# Classification rule with simple select SQL statement

**Spits Warnars H.L.H**

Department of Computing and Mathematics ,Manchester Metropolitan University

John Dalton Building, Chester Street, Manchester M1 5GD, UK

Telp. +44 (0)161 247 1779, fax: +44 (0)161 247 6831

s.warnars@mmu.ac.uk

**Abstract**

A simple sql statement can be used to search learning or rule in relational database for data mining purposes particularly for classification rule. With just only one simple sql statement, characteristic and classification rule can be created simultaneously. Collaboration sql statement with any other application software will increase the ability for creating t-weight as measurement the typicality of each record in the characteristic rule and d-weight as measurement the discriminating behavior of the learned classification/discriminant rule, specifically for further generalization in characteristic rule. Handling concept hierarchy into tables based on concept tree will influence for the successful simple sql statement and by knowing the right standard knowledge to transform each of concept tree in concept hierarchy into one table as to transform concept hierarchy into table, the simple sql statement can be run properly.

1. INTRODUCTION

The attribute oriented induction method has been implemented in data mining system prototype called DBMINER [Han et al. 1996,1997] which previously called DBLearn [Han et al. 1994, 1995a] and been tested successfully against large relational database and datawarehouse for multidimensional purposes. Attribute oriented induction approach is developed for learning different kinds of knowledge rules such as characteristic rules, discrimination or classification rules, quantitative rules, data evolution regularities [Han et al. 1995b], qualitative rules [Han et al. 1993], association rules and cluster description rules [Han and Fu 1995]. Attribute oriented induction has concept hierarchy as an advantage where concept hierarchy as a background knowledge which can be provided by knowledge engineers or domain experts [Han et al. 1992; Han, 1994; Han and Fu, 1995].

Relational database as resources for data mining for mining rules with attribute oriented induction can be read with data manipulation language select sql statement [Meo et al. 1998; Muyeba and Keane, 1999; Zaiane, 2001; Muyeba and Mamadapali, 2005]. Using query for building rules presents efficient mechanism for understanding the mined rules [Imielinski and Virmani, 1999; Muyeba, 2005]. Retrieve records from relational database with select sql statement is known but how to get and implement the simple select sql statement as to implement attribute oriented induction with simple select sql statement as easy and quick to get data result as the understanding.

From database we can learn 2 learning they are:
1)  Positive learning as target class where the data are tuples in the database consistent with the learning concepts

2) Negative learning as contrasting class in which the data do not belong to the target class.

Thus positive learning/target class will be built when do characteristic rule and negative learning/contrasting class will be built when do classification rule where positive learning/target class as result of characteristic rule must be done previously.

Using threshold as a control for maximum number of tuples of the target class in the final generalized relation will not need anymore and as replacement group by operator in sql select statement will limit the final result generalization. Setting different threshold will generate different generalized tuples as the needed of global picture of induction repeatedly as time-consuming and tedious work [Wu et al. 2009]. All interesting generalized tuples as multiple rule can be generated as the global picture of induction by using group by operator or distinct function in sql select statement.

For making easy the implementation a concept hierarchy will just only based on non rule based concept hierarchy and just learning for classification/discriminant rule.

## 2. DATABASE DESIGN

As the connectivity with current or last research data example will refer to data in Cai [1989] and Han et al. [1992] as a concept hierarchy in figure 2 and example of data student in table 1. Figure 1 is the logical data model for database implementation where there are 5 tables, where table student as representative data from table1 dan other tables like Hierarchy_major, Hierarchy_Cat, Hierarchy_GPA and Hierarchy_Birth as the implementation from concept hierarchy in figure 2. Each concept tree from concept hierarchy will be transformed become a table. Database design in figure 1 is similar like star schema in Data Warehouse where table student as fact table and other tables as dimensional table. As a result multi dimensional concept in Data Warehouse can be applied where data can be roll up and drill down and data can be viewed in multiple dimensions with concept slice, dice and pivot [Chen et al. 1996; Han et al. 1999; Cheung et al. 2000). Using aggregate count function and Group by operator in sql select statement will represent the roll up process [Gray et al.1997; Alves and Belo, 2007].

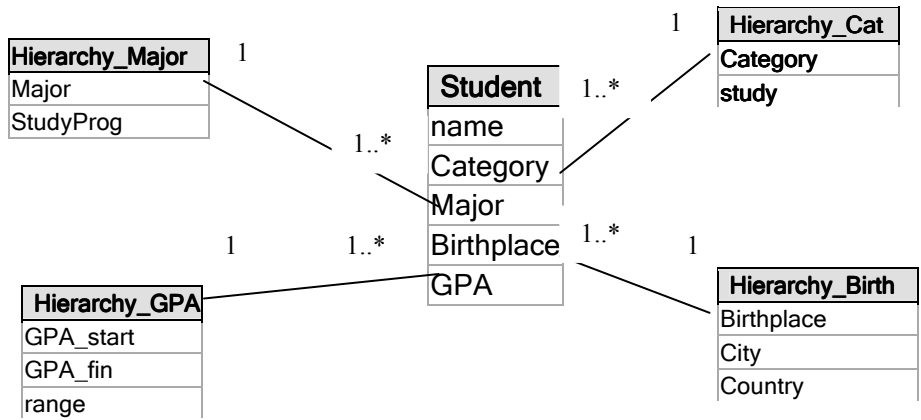

Fig. 1. Logical data model

Table I. Table Student

| Name | Category | Major | Birthplace | GPA |
|------|----------|-------|------------|-----|
| Anton | M.A. | History | Vancouver | 3.5 |
| Andi | Junior | Math | Calgary | 3.7 |
| Amin | Junior | Liberal arts | Edmonton | 2.6 |
| Anil | M.S. | Physics | Ottawa | 3.9 |

| Ayin | Ph.D. | Math | Bombay | 3.3 |
| Amir | Sophomore | Chemistry | Richmond | 2.7 |
| Acai | Senior | Computing | Victoria | 3.5 |
| Abdi | Ph.D. | Biology | Shanghai | 3.4 |
| Afun | Sophomore | Music | Burnaby | 3.0 |
| Agung | Ph.D. | Computing | Victoria | 3.8 |
| Ahing | M.S. | Statistics | Nanjing | 3.2 |
| Akuan | Freshman | literature | Toronto | 3.9 |

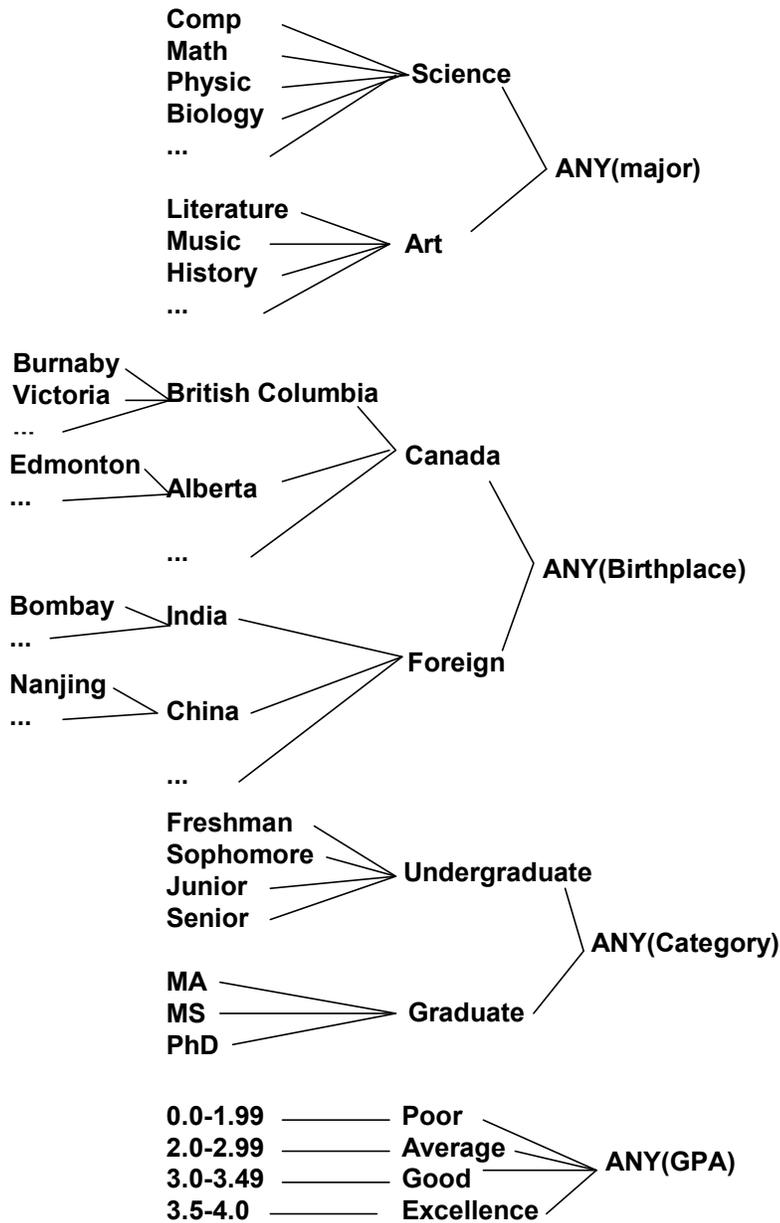

Fig. 2. A concept hierarchy table

## 3. CLASSIFICATION/DISCRIMINATION RULE

Classification/discriminant rule will be done based on data example in Cai [1989] and Han et al. [1992]. The next sql statement will produce table 2 based on generalization strategy steps.

```
select b.study,c.studyprog,d.city,e.range,count(*) as Vote
from student a, hierarchy_cat b, hierarchy_major c, hierarchy_birth d,
hierarchy_gpa e
where a.category=b.category and
      a.major=c.major and a.birthplace=d.birthplace and
      a.gpa>=e.gpa_start and a.gpa<=e.gpa_fin
group by b.study,c.studyprog,d.city,e.range
```

Table II. Classification rule for student category

| study | studyprog | city | Range | Vote |
|---|---|---|---|---|
| graduate | Art | British Columbia | Excellent | 1 |
| graduate | Science | British Columbia | Excellent | 1 |
| graduate | Science | China | Good | 2 |
| graduate | Science | India | Good | 1 |
| graduate | Science | Ontario | Excellent | 1 |
| undergraduate | Art | Alberta | Average | 1 |
| undergraduate | Art | British Columbia | Good | 1 |
| undergraduate | Art | Ontario | Excellent | 1 |
| undergraduate | Science | Alberta | Excellent | 1 |
| undergraduate | Science | British Columbia | Average | 1 |
| undergraduate | Science | British Columbia | Excellent | 1 |

Record 1 until 5 is positive learning/target class and the last records 6 until 11 as negative learning/contrasting class. There are overlapping for record 2 and 11 and as generalization strategy step 8 then the records must be eliminated. The next sql statement will produce table 3.

```
select b.study,c.studyprog,d.country,e.range, count(*) as Vote
from student a, hierarchy_cat b, hierarchy_major c, hierarchy_birth d,
hierarchy_gpa e
where a.category=b.category and
      a.major=c.major and a.birthplace=d.birthplace and
      a.gpa>=e.gpa_start and a.gpa<=e.gpa_fin
group by b.study,c.studyprog,d.country,e.range
```

Table III. Final result classification rule for student category

| Study | studyprog | City | range | Vote |
|---|---|---|---|---|
| Graduate | Art | Canada | Excellent | 1 |
| Graduate | Science | Canada | Excellent | 2 |
| Graduate | Science | Foreign | Good | 3 |
| undergraduate | Art | Canada | Average | 1 |
| undergraduate | Art | Canada | Good | 1 |
| undergraduate | Art | Canada | Excellent | 1 |
| undergraduate | Science | Canada | Excellent | 2 |
| undergraduate | Science | Canada | Average | 1 |

Table 4 is final result with adding t-weight measurement the typicality of each tuple in the characteristic rule and d-weight as measurement the discriminating behavior of the learned classification rule which can be created with application software.

Table IV. Final result classification rule for student category

| Study | studyprog | City | range | Vote | t-weight | d-weight |
|---|---|---|---|---|---|---|
| Graduate | Art | Canada | Excellent | 1 | 1/(1+2+3)=16.67% | 1/(1+1)=50% |
| Graduate | Science | Canada | Excellent | 2 | 2/(1+2+3)=33.33% | 2/(2+2)=50% |
| Graduate | Science | Foreign | Good | 3 | 3/(1+2+3)=50% | 3/3=100% |
| undergraduate | Art | Canada | Average | 1 | 1/(1+1+1+2+1)=16.67% | 1/1=100% |
| undergraduate | Art | Canada | Good | 1 | 1/(1+1+1+2+1)=16.67% | 1/1=100% |
| undergraduate | Art | Canada | Excellent | 1 | 1/(1+1+1+2+1)=16.67% | 1/(1+1)=50% |
| undergraduate | Science | Canada | Excellent | 2 | 2/(1+1+1+2+1)=33.33% | 2/(2+2)=50% |
| undergraduate | Science | Canada | Average | 1 | 1/(1+1+1+2+1)=16.67% | 1/1=100% |

Rule (1) is logical formula for graduate student and rule (2) as logical formula for undergraduate student which can be created with application software.

(1) V(x)=graduate(x)→
(Major(x) Є Art Λ Birthplace(x) Є Canada Λ GPA(x) Є Excellent) [50%]  V
(Major(x) Є science Λ Birthplace(x) Є Canada Λ GPA(x) Є Excellent) [50%]  V
(Major(x) Є science Λ Birthplace(x) Є Foreign Λ GPA(x) Є Good) [100%]

(2) V(x)=undergraduate(x)→
(Major(x) Є Art Λ Birthplace(x) Є Canada Λ GPA(x) Є Average) [100%] V
(Major(x) Є Art Λ Birthplace(x) Є Canada Λ GPA(x) Є Good) [100%]  V
(Major(x) Є Art Λ Birthplace(x) Є Canada Λ GPA(x) Є Excellent) [50%]  V
(Major(x) Є science Λ Birthplace(x) Є Canada Λ GPA(x) Є Excellent) [50%]  V
(Major(x) Є science Λ Birthplace(x) Є Foreign Λ GPA(x) Є Average) [100%]

4. CONCLUSION

By using sql statement for producing table 3 we can produce characteristic rule for graduate student, characteristic rule for undergraduate student and classification/discriminant rule for student simultaneously. Particularly for characteristic rule for further generalization in order to make the last characteristic rule result then performance application software will be needed, specifically for make t-weight, d-weight and logical formula as rules.

Sql statement is the shortest, easy and simple way to get characteristic and classification/discriminant rule from relational database. The powerful sql statement will be increased with application software helping by doing any others sql statement can not be done.

Database design for concept hierarchy as a part of attribute oriented induction will influence the process for sql statement. The knowledge for transformation concept hierarchy will be needed as a basic foundation to do the best select sql statement implementation by transform each of concept tree in concept hierarchy become a table for searching characteristic or classification/discriminant rule from relational database as data mining process.